# On the Composition of Scientific Abstracts

Iana Atanassova[1], Marc Bertin[2] and Vincent Lariviere[3]

[1] iana.atanassova@univ-fcomte.fr
Centre de Recherche en Linguistique et Traitement Automatique des Langues "Lucien Tesnière", Université de Franche-Comté, 30 rue Mégevand, Besançon 25000 (France)

[2] bertin.marc@gmail.com
Centre Interuniversitaire de Recherche sur la Science et la Technologie (CIRST), Université du Quebec à Montréal, CP 8888, Succ. Centre-Ville, Montreal, QC. H3C 3P8 (Canada)

[3] vincent.lariviere@umontreal.ca
École de bibliothéconomie et des sciences de l'information, Université de Montréal, C.P. 6128, Succ. Centre-Ville, Montréal, QC. H3C 3J7 (Canada) and Observatoire des Sciences et des Technologies (OST), Centre Interuniversitaire de Recherche sur la Science et la Technologie (CIRST), Université du Quebec à Montréal, CP 8888, Succ. Centre-Ville, Montreal, QC. H3C 3P8 (Canada)

**Abstract**
Scientific abstracts contain what is considered by the author(s) as information that best describe documents' content. They represent a compressed view of the informational content of a document and allow readers to evaluate the relevance of the document to a particular information need. However, little is known on their composition. This paper contributes to the understanding of the structure of abstracts, by comparing similarity between scientific abstracts and the text content of research articles. More specifically, using sentence-based similarity metrics, we quantify the phenomenon of text re-use in abstracts and examine the positions of the sentences that are similar to sentences in abstracts in the IMRaD structure (Introduction, Methods, Results and Discussion), using a corpus of over 85,000 research articles published in the seven PLOS journals. We provide evidence that 84% of abstract have at least one sentence in common with the body of the article. Our results also show that the sections of the paper from which abstract sentence are taken are invariant across the PLOS journals, with sentences mainly coming from the beginning of the introduction and the end of the conclusion.

**Introduction**

Scientific abstracts contain what is considered by the author(s) as information that best describe documents' content. They represent a compressed view of the informational content of a document and allow readers to evaluate the relevance of the document to a particular information need. According to Hartley (2008), an abstract gives a summary of the content of an article that is comparable to its title and key words but provides different degree of detail: "All articles begin with a title. Most include an abstract. Several include 'key words'. All three of these features describe an article's content in varying degrees of detail and abstraction. The title is designed to stimulate the reader's interest. The abstract summarises the content." (Hartley, 2008: p. 23).

Given the difficulties in obtaining and processing the full-text of scientific documents, as well as the fact that large-scale databases typically index abstracts, most bibliometrics studies use abstracts as a proxy for the content of scientific articles. The motivations for working with abstracts rather than the entire text body of articles are related to the fact that, by definition, abstracts are intended to represent as much as possible the quantitative and qualitative information in documents. Moreover, abstracts are relatively short–between 150 and 300 words—which allows efficient processing and are often available as part of the metadata of scientific articles.

However, abstracts reproduce only part of the information and the complexity of argumentation in a scientific article. Previous work on the topic has provided recommendations on how to write an efficient abstract (Andrade, 2011), on conventions in abstract writing (Hernon & Schwartz, 2010; Swales & Feak, 2009), as well as on the advantages of structured abstracts (Hartley, 2014; Hartley & Sydes, 1997). An important question arises: to what extent and with what accuracy do scientific abstracts reflect article's content? Studying the properties of abstracts and, more specifically, the relationships that exist between abstracts and the full-text of papers can provide important insight into the structure of scientific writing and the possible biases related to representing scientific articles by their abstracts.

Since abstracts include very limited information of an article, they convey only part of the originality and the relevance of the research study. This problem has already been studied by introducing measures of the quality of abstracts (Narine, Yee, Einarson, & Ilersich 1991; Timmer, Sutherland & Hilsden 2003). Other studies focus on the rhetorical structure of scientific abstracts. For example Hirohata, Okazaki, & Ananiadou (2008) proposed a method for the automatic identification of the sections in abstracts using machine learning techniques. Guo, Korhonen, & Liakata, (2010) compared types of categories that appear in abstracts and in the body of articles, and used machine-learning techniques to assign categories to sentences in abstracts independently of the article body, using features such as the position of the sentence, its lexical content and its grammatical structure. Other studies compared scientific abstracts to citation summaries (Elkiss et al., 2008) using metrics based on the weighed cosine similarity, and show that information in citation summaries partly overlaps with abstracts, and citation summaries might contain additional aspects of the paper which are not in the abstract.

*Research question*

The term abstract comes from the Latin verb *abstrahō* that means *"to draw away from, drag or pull away"*. Authors are free, when writing an abstract, to re-use or paraphrase some sentences from the body of their article. The objectives of this paper are, on the one hand, to quantify the re-use of text from the body of the articles in the abstracts, and on the other hand, to identify the zones in the structure of scientific articles that are most likely to contain text that is re-used in the abstract. Working at the level of sentences, which allows us to divide articles into discrete units, we seek to answer the following questions:

1. What percentage of sentences in abstracts are obtained by either direct re-use or a close reformulation of sentences in the body of the papers?
2. Considering the rhetorical structure of the articles, where are located the sentences that serve as sources to produce the abstracts?

Our aim is, thus, to measure the similarity between sentences that appear in abstracts and sentences that are found in the body of articles. Locating the zones in a paper that are used as sources for constructing the abstract, either by direct re-use of their sentences or by reformulations, will give us a better understanding on the parts of an article that are considered as most important by the authors and that, according to them, cover the key elements of the text. If we presume that there exists a stable pattern in writing an abstract, this pattern can be further used in other tasks such as information retrieval or automatic summarization, where the process of filtering out most relevant parts of the text is crucial for obtaining a better document representation. However, if abstracts are mostly made of original sentences, it suggests that they are the result of a human summarization process, where the main ideas of the article have been expressed in a condensed manner making use of novel textual elements.

This study has two limitations. Firstly, in this approach we do not take into account the use of synonyms and other possible reformulation strategies when writing an abstract. Hence, text reuse is likely to be more important than what is estimated in this paper. Secondly, the sample data covers mainly biomedical sciences—except for PLOS ONE which is a multidisciplinary journal—and, hence our results might not be observed in the same manner in other disciplines.

**Methods**

*Dataset*

In order to study the relationships between the full-text of papers and their abstracts, we processed a large collection of research articles. The dataset we used consists of all articles published by the seven peer-reviewed journals of the Public Library of Science[1] (PLOS): *PLOS Biology, PLOS Computational Biology, PLOS Genetics, PLOS Medicine, PLOS Neglected Tropical Diseases, and PLOS Pathogens* and *PLOS ONE*, a journal that covers all fields of science and social sciences. These seven journals follow the same publication template, where authors are explicitly encouraged to use the IMRaD structure. Our dataset contain all articles published up to September 2013. The articles are accessible from the publisher in XML format as structured full text. The content of the articles is represented using the Journal Article Tag Suite[2] (JATS), where the abstract is present as a separate XML element which is part of the metadata, and the textual content of the article is given in the *body* element, which is further divided into sections and paragraphs. The author guidelines for research articles in PLOS journals require that each article contain an abstract of one paragraph limited to 300 words, except for *PLOS Biology* and *PLOS Medicine* that do not have word limit for abstracts. PLOS defines the abstract as follows[3]

> *"The abstract succinctly introduces the paper. It should mention the techniques used without going into methodological detail and mention the most important results. The abstract is conceptually divided into the following three sections: Background, Methodology/Principal Findings, and Conclusions/Significance. However, the abstract should be written as a single paragraph without these headers. Do not include any citations in the abstract. Avoid specialist abbreviations."*

An author summary of 150-200 words is included in all research articles, except for publications in *PLOS ONE* and *PLOS Medicine*. It should provide a non-technical summary of the work and it should be distinct from the scientific abstract. The guidelines for writing author summaries are as follows[4]:

> *"Distinct from the scientific abstract, the author summary should highlight where the work fits in a broader context of life science knowledge and why these findings are important to an audience that includes both scientists and non-scientists. Ideally*

---

[1] http://www.plos.org/
[2] http://jats.nlm.nih.gov/
[3] http://journals.plos.org/plosbiology/s/submission-guidelines (accessed June, 2015)
[4] http://journals.plos.org/plosntds/s/submission-guidelines (accessed June, 2015)

> *aimed to a level of understanding of an undergraduate student, the significance of the work should be presented simply, objectively, and without exaggeration."*

Our study focuses mainly on the abstracts of research articles. However, for the sake of comparison, we will examine also some properties of the author summaries. Table 1 presents the number of articles for each journal, as well as the mean article length, the mean abstract length and the mean author summary length, expressed as number of sentences.

Table 1. Characteristics of the PLOS dataset

| Journal | Nb articles | Nb author summaries | Avg article length | Avg abstract length | Avg author summary length |
|---|---|---|---|---|---|
| PLOS Biology | 1,754 | 1,171 | 231.298 | 8.837 | 8.000 |
| PLOS Computational Biology | 2,560 | 2,337 | 259.037 | 9.244 | 7.600 |
| PLOS Genetics | 3,414 | 3,158 | 231.242 | 9.313 | 7.693 |
| PLOS Medicine | 926 | 0 | 168.272 | 13.029 | |
| PLOS Neglected Tropical Diseases | 1,872 | 1,863 | 171.643 | 11.444 | 7.982 |
| PLOS Pathogens | 2,976 | 2,837 | 234.309 | 9.456 | 7.768 |
| PLOS ONE | 72,158 | 0 | 177.657 | 9.935 | |
| Total | 85,660 | 11,366 | 185.059 | 9.917 | 7.772 |

*Segmentation and section titles processing*

PLOS author guidelines encourage authors to use the IMRaD structure for research articles. While most articles in the corpus contain all four sections (Introduction, Methods, Results and Discussion), the order in which these sections appear can vary. Similarly, while PLOS requires that the argumentative structure of articles follows this specific pattern, slight variations are possible in the section titling. For example, the Methods section can be named *"Materials and Methods"* or *"Methods and Model"*. In order to categorize the sections we had to take into account such variations. This approach has been described in Bertin, Atanassova, Larivière, & Gingras (2015). Table 2 presents the number and percentage of research articles that contain all four section types of the IMRaD structure. It shows that almost 98% of the articles in the corpus contain the four section types, and for all journals but PLOS Computational Biology, this percentage is greater than 98%.

Table 2. Research articles that contain the four section types of the IMRaD structure

| Journal | Articles that contain all four section types | Percentage |
|---|---|---|
| PLOS Biology | 1,735 | 98.92% |
| PLOS Computational Biology | 2,418 | 94.45% |
| PLOS Genetics | 3,402 | 99.65% |
| PLOS Medicine | 915 | 98.81% |
| PLOS Neglected Tropical Diseases | 1,867 | 99.73% |
| PLOS Pathogens | 2,973 | 99.90% |
| PLOS ONE | 70,583 | 97.82% |
| Total | 83,893 | 97.94% |

*Similarity measures*

In order to assess the similarity between abstracts and the body of articles, we segmented article bodies, abstracts and author summaries into sentences. In general, the similarity measures applied to a pair of texts assign a similarity score between 0 and 1 which expresses to what extent the first text segment resembles the second in terms of the number of common words or collocations. A similarity of 0 means that the text segments are completely different, while a similarity of 1 means that the texts are identical. For our task, we have used a combination of three similarity measures, that come from character-based and term-based similarity measures' approaches. The similarity measures are defined as follows:

1. **Exact Substrings**. We consider two segments as similar, if one of the segments is an exact substring of the other. The similarity measure is calculated as follows:

$$SIM_E(A,B) = \begin{cases} 1, & \text{if } A \text{ is substring of } B \text{ or } B \text{ is substring of } A \\ 0, & \text{otherwise.} \end{cases}$$

2. **Cosine Similarity.** Cosine similarity is one of the most popular similarity measures for text documents and has been applied in numerous studies in information retrieval (Salton & Buckley, 1988). It measures the cosine of the angle between two vectors. We represented text segments as term vectors, where stop-words were cleared using WEKA (Hall, Frank, & Holmes, 2009). If *A* and *B* are m-dimensional vectors over the term set $\{t_1, ..., t_m\}$, then their cosine similarity is:

$$SIM_C(A,B) = \frac{\sum_{i=1}^{m} a_i \cdot b_i}{\sqrt{\sum_{i=1}^{m} a_i^2} \sqrt{\sum_{i=1}^{m} b_i^2}}.$$

Cosine similarity is bound in the interval [0,1]. If it is 1, this means that the two documents are represented by the same vectors after normalization.

3. **Levenshtein Distance.** In information theory and computer science, the Levenshtein distance *Lev(A,B)* is a string metric which measures edit distance (Levenshtein, 1966).

$$SIM_L(A,B) = 1 - \frac{Lev(A,B)}{max(|A|,|B|)}$$

We have considered the term-level Levenshtein distance between sentences, which is given by the minimum number of operations, which are needed to transform one sentence into the other, where an operation is an insertion, a deletion, or a substitution of a term. The Levenstein similarity measure is calculated as follows :

There exist a large number of other similarity measures, for example Jaro-Winkle, Smith-Waterman, N-gram, as well as corpus-based similarities (Gomaa & Fahmy, 2013). In this first study on the relation between abstracts and the body of articles, we have chosen to work with the three similarity measures that we have defined above and that are among the most widely used similarity measures for text processing. Apart from these three measures we have also performed the calculations using other term-based similarity measures, namely Dice's coefficient and Jaccard similarity. The results obtained by the use of these two measures being almost identical to the results of the Cosine Similarity, we report only the latter in this article for the sake of concision.

If we consider that the abstract of an article contains the set of sentences *{A₁, ..., Aₙ}*, to measure the similarity between a sentence and the article's abstract, for each sentence *S* in the body we calculate the score *SIM$_E$(S)*, *SIM$_C$(S)* and *SIM$_L$(S)* which is the maximum of the similarities between *S* and the set *{A₁, ..., Aₙ}*.

$$SIM_J(S) = \max_{i=1}^{n} (SIM_J(S, A_i)), \text{ where } J \in \{E, C, L\}$$

For the following experiment we will consider that a sentence $A_k$ from the abstract matches a sentence *S* from the body of an article if any of the three similarity measures is above a threshold T that we fix at T=0.6:

$$SIM_J(S, A_k) = SIM_J(S) \geq 0.6 \text{ where } J \in \{E, C, L\}$$

We define the overall similarity of a sentence in the $A_k$ abstract to the body of the article as the maximal similarity between $A_k$ and the sentences in the body:

$$SIM_{max}(A_k) = \max_{J \in \{E,C,L\}, S} (SIM_J(S, A_k))$$

**Results**

We analyse the similarities between sentences in abstracts and article bodies according to three different criteria:

1. the percentage of sentences in abstracts that present a high similarity with sentences found in the body of the article;
2. the position of sentences along the IMRaD structure that are also used in the abstracts;

As the corpus contains both abstracts and author summaries, we will first study the differences between them.

*Abstracts and Author Summaries Text Re-use*

Table 3 presents the percentage of sentences in abstracts and author summaries having similarities with sentences in article bodies of 1, between 1 and 0.8, between 0.8 and 0.6, and below 0.6. The table gives the percentages for each of the three similarity measures and the last line is obtained by calculating the maximum of the three similarities for each sentence in the abstracts and author summaries.

**Table 3. Percentages of sentences in abstracts and author summaries that match sentences from the body of the articles**

|  | SIM (S, A$_k$) = 1 | 1 > SIM (S, A$_k$) ≥ 0.8 | 0.8 > SIM (S, A$_k$) ≥ 0.6 | 0.6 > SIM (S, A$_k$) |
|---|---|---|---|---|
| **In abstracts** |  |  |  |  |
| SIM$_E$ | 1.66% |  |  | 98.34% |
| SIM$_C$ | 1.06% | 4.53% | 16.91% | 77.50% |
| SIM$_L$ | 0.64% | 1.40% | 2.92% | 95.03% |
| SIM$_{max}$ | 2.02% | 4.80% | 16.93% | 76.26% |
| **In author summaries** |  |  |  |  |
| SIM$_E$ | 0.70% |  |  | 99.30% |
| SIM$_C$ | 0.66% | 1.97% | 8.82% | 88.56% |
| SIM$_L$ | 0.36% | 0.78% | 1.60% | 97.26% |
| SIM$_{max}$ | 0.90% | 2.53% | 10.44% | 86.13% |

The sum of the first three columns shows that more than 23% of all sentences in abstracts have similarities above 0.6 with sentences found in the article body. This first result quantifies text re-use in scientific abstracts. The table also shows that this phenomenon is less present in author summaries, which contain only about 12% of sentences that match sentences in the article body. The editorial requirements limit both author summaries and abstracts to 300 words. We have examined the lengths of author summaries and abstracts in terms of number of sentences. We note that the lengths of the sentences in author summaries and in abstracts are very close: 23.35 words on average for author summaries and 23.55 words on average for abstracts.

Figure 1 presents the distribution of the abstract and author summary lengths in the corpus in terms of number of sentences. The horizontal axis gives lengths as number of sentences and the vertical axis gives the percentage of abstracts and author summaries. The mean values are indicated by the vertical dashed lines. The large majority of abstracts are composed of 7 to 13 sentences, while author summaries tend to be shorter with a mean around 8 sentences. The figure also shows that abstract lengths are relatively variable with about 20% of abstracts having less than 7 sentences and another 20% having more than 13 sentences. As for author summaries, the vast majority, more than 75%, have between 7 and 10 sentences. For the following analyses we concentrate mainly on abstracts.

**Preprint:** I. Atanassova, M. Bertin, V. Lariviere (2016) On the Composition of Scientific Abstracts, *Journal of Documentation*, vol. 72, issue 4. Submitted 12-Sep-2015, accepted 10-Feb-2016.

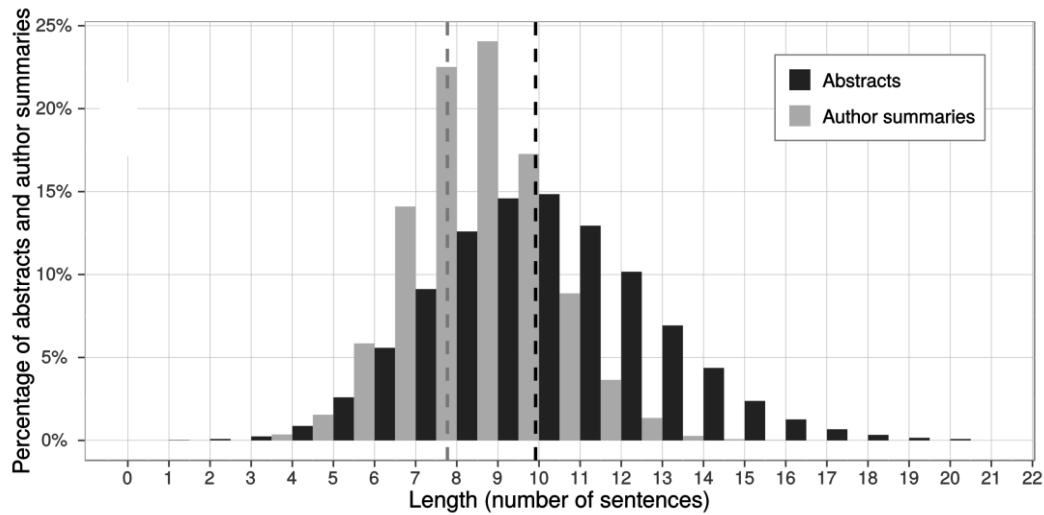

**Figure 1. Abstract and author summary length distribution**

*Text Re-use by Journal*

We characterize the differences, in the seven PLOS journals, in abstracts re-use text from the body of the articles in the seven journals in the corpus. Figure 2 presents the overall percentage of sentences that have a maximal similarity above 0.8 and above 0.6 with sentences from the article body. This figure shows some major differences across journals. On the one hand, abstracts from *PLOS Neglected Tropical Diseases* and *PLOS ONE* have a very high percentage of sentences that are re-used from the article body. In *PLOS ONE*, more than 25% of sentences in abstracts are very similar to sentences in the article body. For the five other PLOS journals, the percentage of similar sentences is between 12.5% and 15.1%.

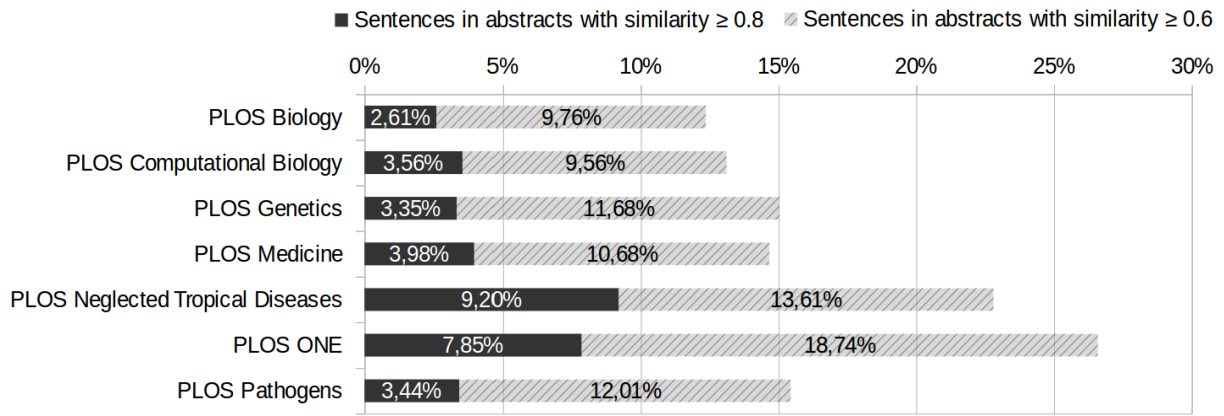

**Figure 2. Percentage of sentences in abstracts with similarity above 0.8 and 0.6 in the seven journals**

The number of sentences in each abstract that are similar to sentences in article body varies among the journals and among the articles of the journals. Table 4 presents the percentage of abstracts in each journal with 0 sentence, 1 sentence, 2 or 3 sentences, and more than 3 sentences that are very similar to other sentences found in the body of the article. The first column shows that around 16% of abstracts are composed entirely of original sentences that are not similar with any sentence in the article body. The remaining 84% of abstracts contain at least one sentence similar to a sentence in the article body and more than 33% of abstracts contain more than 3 such sentences. We can observe that *PLOS Medicine* and *PLOS Neglected Tropical Diseases* are characterized by a very high number of abstracts having more than 3 sentences that match sentences in article bodies.

Table 4. Text re-use in abstracts for the seven PLOS journals: by number of sentences

| Journal | 0 sentences | 1 sentence | 2 or 3 sentences | More than 3 sentences |
|---|---|---|---|---|
| PLOS Biology | 26.74% | 23.38% | 31.93% | 17.96% |
| PLOS Computational Biology | 24.18% | 23.98% | 28.71% | 23.13% |
| PLOS Genetics | 17.40% | 21.00% | 31.40% | 30.20% |
| PLOS Medicine | 4.43% | 7.67% | 18.57% | 69.33% |
| PLOS Neglected Tropical Diseases | 9.24% | 14.64% | 27.94% | 48.18% |
| PLOS ONE | 15.77% | 18.72% | 31.78% | 33.73% |
| PLOS Pathogens | 15.76% | 18.55% | 32.19% | 33.50% |
| *Total* | *16.04%* | *18.85%* | *31.47%* | *33.64%* |

As abstracts vary in length, we have also examined the relative proportion of each abstract that comprises sentences similar to those found sentences in the article body. Table 5 presents the percentage of abstracts in each journal composed of up to 25%, 50%, 75% and 100% of sentences similar to those of the article body. The main journal, *PLOS ONE*, presents very high percentages of text re-use: in almost 17% of its papers, more than half of abstracts' sentences have a very high level of similarity with sentences of the article body. For the other six journals, text re-use is less important. The first two columns show that the vast majority of abstracts in these journals (a total of 55.87%) are composed of less than 25% of sentences similar to sentences found in the article body.

Table 5. Text re-use in abstracts for the seven PLOS journals: by percentage of abstracts' text

| Journal | No text re-use | 0%-25% text re-use | 25%-50% text re-use | 50%-75% text re-use | 75%-100% text re-use |
|---|---|---|---|---|---|
| PLOS Biology | 26.74% | 57.47% | 14.08% | 1.65% | 0.06% |
| PLOS Computational Biology | 24.18% | 59.10% | 14.41% | 2.23% | 0.08% |
| PLOS Genetics | 17.40% | 61.31% | 19.19% | 2.02% | 0.09% |
| PLOS Medicine | 4.43% | 78.40% | 16.09% | 1.08% | 0.00% |
| PLOS Neglected Tropical Diseases | 9.24% | 59.13% | 27.51% | 3.74% | 0.37% |
| PLOS ONE | 15.77% | 35.79% | 31.68% | 13.91% | 2.86% |
| PLOS Pathogens | 15.76% | 62.13% | 19.76% | 2.32% | 0.03% |
| *Total* | *16.04%* | *39.83%* | *29.63%* | *12.07%* | *2.42%* |

*Location of re-used sentences throughout the IMRaD Structure*

As we have shown on Table 3, more than 23% of sentences in abstracts are similar to sentences in article body. Here, we study the position of these sentences in the structure of the articles in order to reveal which rhetorical zones contain the most important information from the point of view of the authors, which increases the likelihood of intertextuality. Table 6 presents the percentage of sentences in each section type of the IMRaD structure that match sentences in abstracts. The Introduction section contains the highest percentage of such sentences and the Methods section contains the lowest percentage. This is true for all journals except for *PLOS Medicine*, where the Results section displays a higher percentage than the Introduction section.

**Table 6. Percentage of sentences in the four section types that match sentences in abstracts**

| Journal | I | M | R | D | *Total* |
|---|---|---|---|---|---|
| PLOS Biology | 3.23% | 0.08% | 0.71% | 1.57% | 1.03% |
| PLOS Computational Biology | 2.71% | 0.19% | 0.58% | 1.42% | 0.93% |
| PLOS Genetics | 3.67% | 0.14% | 1.07% | 1.88% | 1.31% |
| PLOS Medicine | 5.05% | 1.46% | 5.12% | 3.29% | 3.41% |
| PLOS Neglected Tropical Diseases | 3.88% | 0.86% | 3.49% | 2.83% | 2.53% |
| PLOS ONE | 4.40% | 0.62% | 3.01% | 3.48% | 2.57% |
| PLOS Pathogens | 3.72% | 0.09% | 1.38% | 2.17% | 1.46% |
| *Total* | *4.24%* | *0.56%* | *2.61%* | *3.24%* | *2.36%* |

The last column represents the total percentage of sentences in all four sections that match sentences in the abstract. The journal *PLOS Medicine* stands out as having abstracts that re-use more that 3% of the text, with more than 5% from the Introduction and Results sections. This is due to the fact that articles in *PLOS Medicine* tend to be shorter and abstracts tend to be longer compared to the other journals (see table 1). Figure 3 presents the normalized distribution of sentences in the IMRaD structure that have maximal similarity with sentences in abstracts above 0.6. The horizontal axis represents the text progression from 0% to 100% in the IMRaD structure in terms of number of sentences. The vertical axis gives the average percentage of sentences at a given point of the text for each journal. The vertical lines on the graph indicate the average positions of the sections boundaries. Part of the articles in the corpus contain all four section types but in a different order. To obtain this representation, sections were reordered where necessary to follow the standard order: Introduction, Methods, Results, Discussion. It shows that, in all the journals, the distributions are very similar, which suggests that there exists a strong relation between the rhetorical structure of articles and the zones that authors re-use when writing abstracts. The highest percentage of sentences is located in the beginning of the Introduction and in the end of the Discussion sections, with an important peak in the second part of the Introduction.

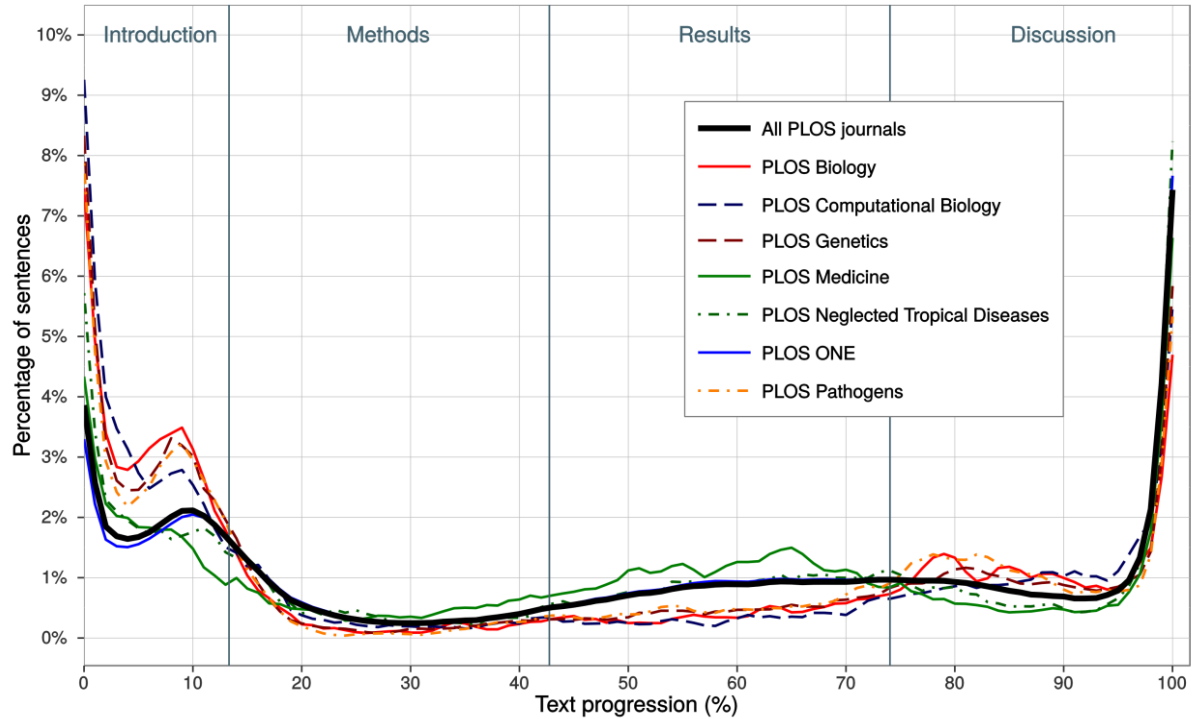

**Figure 3. Distribution of sentences in article body having similarity with sentences in abstracts above 0.6**

**Discussion and Conclusion**

This paper provides a first analysis of the similarity between the text of scientific abstracts and the body of articles, using sentences as the basic textual unit. Our results show that about 16% of abstracts are composed entirely of original sentences and the remaining 84% contain at least one sentence which is similar to a sentence in the body of the article. Overall, an average of 23% of the sentences in abstracts are close reformulations of sentences in the body of articles. The similarity measures that we use in this study allow us to detect only a part of the paraphrases and reformulations that can exist between sentences in abstracts and sentences in the article body.

The curves found in Figure 3 for the seven journals are very similar to each other for the Methods, Results and Discussion sections, which suggests that the specific places in papers where abstract text comes from is, globally, invariant across domains. They also show that that the content of the four sections is represented differently in the abstracts, and that sentences from the Introduction section—and to a lesser extent, the conclusion section—are re-used in abstracts much more often than sentences in the other sections. This suggest that these two sections are considered by the authors as the most representative of the content of the article, much more than the methods and results sections.

Some differences are, however, present in the Introduction section. For example, we can observe that the values for PLOS Biology, PLOS Genetics and PLOS Pathogens are relatively high and present a local maximum at around 9% of the text, while the curve of PLOS Medicine diminishes steadily throughout the Introduction. Considering the curve for all the seven journals, we can define four different zones: zone A from the beginning to the first local minimum around 4%; zone B from the first local minimum to the first local maximum around 9%; zone C from the first local maximum to the start of the increase around 95%; and zone D for the last 5% of the article. These zones in the text, taking into consideration the IMRaD sequence, convey specific types of information in the organization of research articles. Zone A, which is the beginning of the Introduction, typically states the research topics. Zone D contains the last paragraphs of the article which sum up the obtained results. These two zones contain the largest amount of the linguistic material that forms the abstracts.

As the goal of our study was to characterize the relationship between the abstract and the full text of papers rather than performing an exhaustive detection of paraphrases, we did not rely on synonyms and other reformulation strategies that can be used in a scientific abstracts. Despite this limitation, our results do provide new insights for improving automatic abstracting tools as well as information retrieval approaches, in which text organization and structure are important features. Measuring the similarity between sentences, paragraphs in scientific abstract and the body of text is an important component also for document clustering, machine translation, and text summarization. Furthermore, the position of sentences in the body of articles that are re-used in the abstract give important indications on the structure of scientific papers and the relevance of its different parts as perceived by the author.

Further research in this topic should refine these results by introducing lexical and semantic similarity measures. For example, various dictionary-based algorithms allow to capture the semantic similarity between words and sentences (e.g. Banerjee & Pedersen, 2002; Jiang & Conrath, 1997; Gabrilovich & Markovitch, 2007). The application of such algorithms should allow to obtain higher recall in the detection of paraphrases. The results of our study are to be related to the work around the logical structure of abstracts and recommendations for their writing. Indeed, the works of Šauperl, Klasinc & Lužar (2008) and Jamar, Šauperl & Bawden (2014) show that the abstract should follow a structure similar to the IMRaD structure, but that the authors seldom follow such recommendations. In this perspective, our methods could be part of authoring tools for good practices in the writing of abstracts.

**Acknowledgments**
We thank Benoit Macaluso of the Observatoire des sciences et des technologies (OST-UQAM) for harvesting and providing the PLOS dataset.